\documentclass[]{spie}  
\usepackage[]{graphicx}

\title{Gemini planet imager integration to the Gemini South telescope software environment} 

\author{
Fredrik T. Rantakyr\"o\supit{a}, Andrew Cardwell\supit{a}, Jeffrey Chilcote\supit{b}, Jennifer Dunn\supit{c}, Stephen Goodsell\supit{d}, Pascale Hibon\supit{a}, Bruce Macintosh\supit{e}, Carlos Quiroz\supit{a}, Marshall D. Perrin\supit{f}, Naru Sadakuni\supit{a}, Leslie Saddlemyer\supit{c}, Dmitry Savransky\supit{g}, Andrew Serio\supit{a}, Claudia Winge\supit{a}, Ramon Galvez\supit{a}, Gaston Gausachs\supit{a}, Kayla Hardie\supit{a}, Markus Hartung\supit{a}, Javier Luhrs\supit{a}, Lisa Poyneer\supit{h} and Sandrine Thomas\supit{i}
\skiplinehalf
\supit{a}Gemini Observatory, Casilla 603, La Serena, Chile; \\
\supit{b}Department of Physics and Astronomy, UC Los Angeles, Los Angeles, CA 90095 USA; \\
\supit{c}National Research Council Canada, Herzberg Institute of Astrophysics, 5071 West Saanich Road, Victoria, British Columbia V9E 2E7, Canada;\\
\supit{d}Gemini Observatory, 670 N. A'ohoku Place, Hilo, USA; \\
\supit{e}Stanford University, Stanford, California 94305, USA; \\
\supit{f}Space Telescope Institute, 3700 San Martin Drive, Baltimore MD 21218, USA;\\
\supit{g}Cornell University, 144 East Ave., Ithaca, NY 14850, USA;\\
\supit{h}Lawrence Livermore National Lab, 7000 East Ave, Livermore, CA, 94550, USA;\\
\supit{i} NASA Ames Research Center, Moffett Field, CA 94035, USA
}

\authorinfo{Further author information: (Send correspondence to Fredrik T Rantakyr\"o)\\Fredrik T. Rantakyr\": E-mail: frantaky@gemini.edu, Telephone: 56(51) 2204665}

\begin{document} 
  \maketitle 

\begin{abstract}
The Gemini Planet Imager is an extreme AO instrument with an integral field spectrograph (IFS) operating in Y, J, H, and K bands. Both the Gemini telescope and the GPI instrument are very complex systems. Our goal is that the combined telescope and instrument system may be run by one observer operating the instrument, and one operator controlling the telescope and the acquisition of light to the instrument. This requires a smooth integration between the two systems and easily operated control interfaces. We discuss the definition of the software and hardware interfaces, their implementation and testing, and the integration of the instrument with the telescope environment.
\end{abstract}



\section{INTRODUCTION}
\label{sec:intro}  
The Gemini Planet Imager (GPI) is a dedicated facility instrument for direct imaging and spectroscopy, and polarimetric observations of extrasolar planets. It combines a very high-order adaptive optics system, a diffraction-suppressing coronagraph, and an integral field spectrograph with low spectral resolution but high spatial resolution. Every aspect of GPI has been tuned for maximum sensitivity to faint planets near bright stars.

The Gemini Observatory consists of twin 8.1-meter diameter optical/infrared telescopes located on two of the best observing sites on the planet. From their locations on mountains in Hawai'i and Chile, Gemini Observatory's telescopes can collectively access the entire sky. Gemini is operated by a partnership of six countries including the United States, Canada, Chile, Australia, Brazil and Argentina. 

This document describes the process of integrating the GPI instrument to the telescope environment on Gemini South telescope located on Cerro Pachon near La Serena in Chile. 
\section{GPI-Gemini Planet Imager}
 GPI is an extreme adaptive-optics imaging polarimeter/integral-field spectrometer, which provides diffraction-limited data between 0.9 and 2.4 microns. The system provides contrast ratios of 10$^6$ on companions at separations of 0.2-1 arcsecond in a 1-2 hour observation. The science instrument provides spectroscopy or dual-beam polarimetry of any object in the field of view. Bright natural guide stars (I$<$9 mag) are required for optimal performance of the GPI adaptive optics system. GPI will be capable of detecting point sources down to H=20 magnitude, with $\leq$5-sigma, in 1 hour (absent photon noise from a bright companion). GPI combines four main optomechanical systems, each controlled by its own computer system and the totality of the system controlled by the TLC(Top Level Computer, see section,\, \ref{sec:tlc}): 
\begin{itemize}
\item The adaptive optics (AO) system, responsible for fast measurement of the instantaneous wave front, and for providing wave front control via two deformable mirrors.
\item The calibration unit (CAL) is a high-accuracy infrared wave front sensor tightly integrated with the coronagraph. It provides pointing and focus sensing to keep the target star centered on the coronagraph with 1mas accuracy and slow low to high-order aberration corrections.
\item The coronagraph uses a combination of apodized masks and focal-plane stops to control diffraction and pinned speckles.
\item The integral field spectrograph (IFS) produces the final science image, including simultaneous multiple channels to suppress residual speckle noise in the spectroscopy mode or polarimetric imaging allowing the determination of the four Stokes parameters.
\end{itemize}
The primary data product of the IFS is a data cube consisting of slightly more than 190x190 spatial locations, each with typically 18 spectral channels in spectroscopy mode and two spots in each positions for the polarimetric mode. The final field of view (FOV) is 2.7 arcseconds on a side, with 14.3 milliarcsecond sampling.

Further details please read the several papers in this conference\cite{dunnthis,hartungthis,konopackythis,larkinthis,perrinthis,poyneerthis} on the instrument and the general GPI paper by Macintosh, et al.\,2014\cite{macintosh2014first}. 
\subsection{GIAPI-Gemini Instrument Application Programming Interface}
\label{sec:giapi} 
\begin{figure}
   \begin{center}
   \begin{tabular}{c}
   \includegraphics[height=10cm]{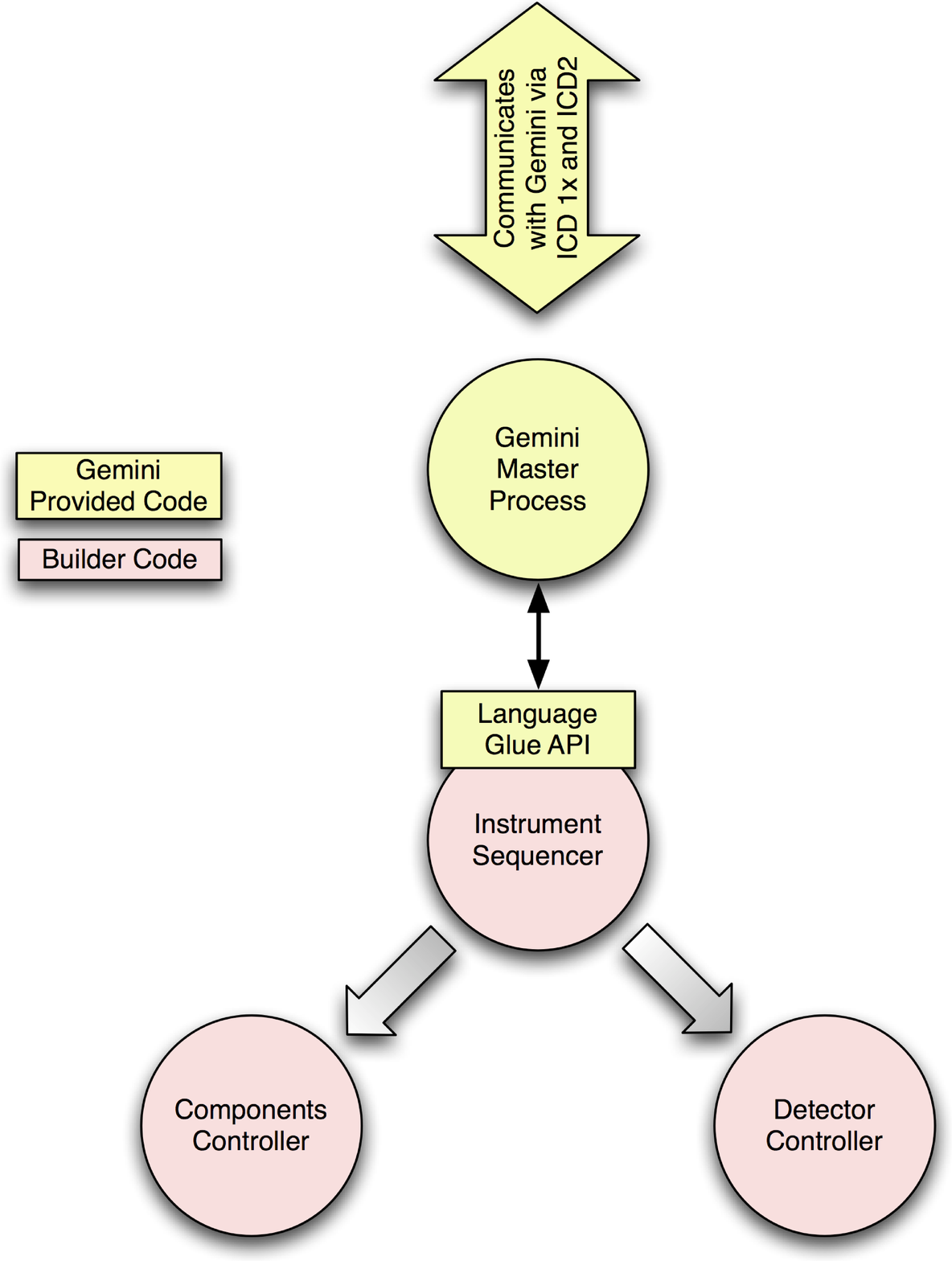}
   \end{tabular}
   \end{center}
   \caption[] 
   { \label{fig:giapi} 
GIAPI based instrument architecture.}
   \end{figure}

The Gemini Instrument Application Programming Interface (GIAPI) is a library and middleware aiming to simplify Instrument Software integrating to Gemini’s Observatory Control System (OCS). Providing a unified library for integration reduces support costs and reduces heterogeneity of a less cohesive approach. The library is currently provided in C++ but could be easily extended to Java and other popular programming languages.
GIAPI provides a minimal set of software services to facilitate instrument’s software integration. These services cover the following tasks:
\begin{itemize}
\item Provide Status and Health Information to Gemini
\item Accept and handle sequencing commands
\item Publish events when datasets are acquired
\item Access to Observatory status and services, like current time, telescope’s state, etc.
\end{itemize}
As part of the GIAPI, there is a middleware component, called the Gemini Master Process (GMP), that lives between the instrument and the OCS. The GMP hides the complexity of interacting with the OCS and allows both sides to evolve more independently. Figure\,\ref{fig:giapi} shows the architecture of a prototypical GIAPI-based instrument.

GPI is the first instrument to use the GIAPI to integrate to the OCS. The integration process was simplified thanks to the use of the GIAPI and let the instrument to be controlled by the OCS very shortly after arriving to Gemini. 
\subsection{OCS-Observatory Control System}
\label{sec:ocs} 
The OCS is all the software that composes the user interface and high level control of the observatory. The system as a whole provides tools that assist the astronomer from the proposal submission phase through planning, observation execution, and data review. These tools are principally the Telescope Control System (TCS), Telescope Component Control (TCC), Observing Data Base (ODB), and, the Sequence Executor (SeqExec).  

The telescope control system (TCC) was created by David Terrett of Rutherford labs in the UK, it is written in tcl/tk. The main purpose of the TCC is to simplify the operations of the telescope on a nightly basis by selecting only those operations of the TCS which are necessary and most commonly used by the system operators. In simplified terms the TCC is the front-end to the TCS, thus we can treat the two systems as a single whole for the purpose of this document. 
\subsection{TLC-Top Level Computer}
\label{sec:tlc}  
\begin{figure}
   \begin{center}
   \begin{tabular}{c}
   \includegraphics[height=7.5cm,width=16cm]{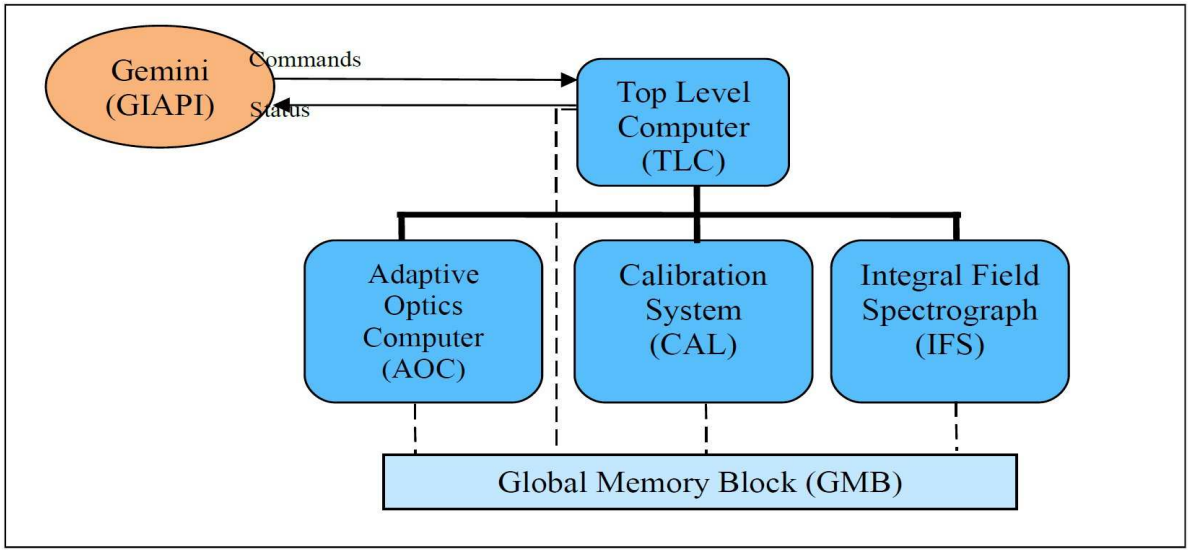}
   \end{tabular}
   \end{center}
   \caption[] 
   { \label{fig:tcs} 
The TLC software structure.}
   \end{figure}

There are 4 software systems that join together to provide the functionality of GPI.  There is the Top Level Computer (TLC) which performs the movement of the slow moving mechanisms, is the main interface for commands and status with Gemini, and internal sequencing of GPI.  The AOC performs the adaptive optics reconstruction, the CAL provides tip/tilt/focus measurements to keep the target star centered, and the IFS is the science instrument which performs the observations.  GPI’s command structure is hierarchical and the TLC is the one that initiates commands on the other software systems.  The client-server architecture is realized using Remote Procedure Calls.  Status information is easily transmitted between machines using the GPI global memory block which transmits changes in to all of the machines. The software architecture is shown in figure\,\ref{fig:tcs}.  

GPI also interfaces with the Gemini Telescope Control System, Primary Mirror Control and Data Handling System.  This is made virtually invisible to the TLC using Gemini’s new interface called the Gemini Application Interface (GIAPI).  
\subsection{GPILIB-The IDL console interface}
\label{sec:idl}  
In order to provide more flexibility in interacting with the instrument, the GPI integration, testing and commissioning teams have developed a scriptable interface to the instrument, written entirely in IDL and collectively called gpilib.  This library of codes can be used to command GPI actions (such as moving mirrors, triggering AO system loops, or taking IFS images), can query and return data from GPI's global memory block, and can be used to easily dump and visualize AO and CAL system telemetry.

Since all of this functionality is available via an IDL command line, this provides a powerful engineering interface to the instrument which can be used to carry out non-routine calibration tasks or rapid debugging if errors arise.  While interaction with the instrument is provided by wrapping pre-compiled binaries that call the same functionality that is executed from the GUI interfaces and called via Gemini's SeqExec, gpilib also contains some specialized routines not available in the rest of GPI's
codebase.  These include tools for visualization and analysis of AO telemetry (including vibration analysis), tools for tuning advanced LQG tip/tilt controllers\cite{lisa.poyneer.spie2014}\,, and routines implementing focal-plane wave front control\cite{dmitry.savransky.spie2012,dmitry.savransky.spie2014}\,.
\section{Requirements and Development}
This process was broken into two major steps. First, writing a Software Requirement document that specified all foreseen software changes to allow smooth integration of GPI into the telescope environment and second taking the requirements document and transforming each of the requirements into detailed software tasks using a JIRA\footnote{https://confluence.atlassian.com/display/JIRA/JIRA+101} software development tool.  
\subsection{Software requirement document}
\label{sec:swreq}  
The first version of the document was completed using the internal GPI builder documentation and internal Gemini documentation as basis to formulate the requirements. This was completed in November 2009 and the document has undergone 34 version updates after this first version. Most of the changes where due to minor changes and added functionality in GPI functionality. The writing of the requirements required many iterations with the instrument builder team and the Gemini system engineering group so that all required functions available with the instrument could be accessed through calls through the GIAPI (section\,\ref{sec:giapi}) layer using the standard Gemini software tools (sections\,\ref{sec:ocs},  \ref{sec:pit} and \ref{sec:seqexec}). 

The final requirements can be broken up into:
\begin{itemize}
\item General {\it Software Functional Requirements}: 12 requirements
\item {\it Observing Tool(OT)}: 55 requirements
\item {\it OT Position Editor}: 4 requirements 
\item {\it Overheads}: 3 requirements
\item {\it Seqexec}: 46 requirements
\item {\it TCC/TCS Requirements}: 15 requirements
\item {\it WDBA}(internal Gemini instrument configuration database): 2 requirements  
\item {\it Electronic Observing log}, 1 requirement
\item {\it Queue Planning Tool}: 2 requirements
\item {\it PIT}: 18 requirements
\item {\it Instrument Status Display(ISD)}: 7 requirements
\item {\it Gemini Weather System(GWS)}: 1 requirement
\item {\it WFS Diagnostic Displays}: 14 requirements
\item {\it Database archival}: 2 requirements
\end{itemize}

   \begin{figure}
   \begin{center}
   \begin{tabular}{c}
   \includegraphics[height=8cm, width=16cm]{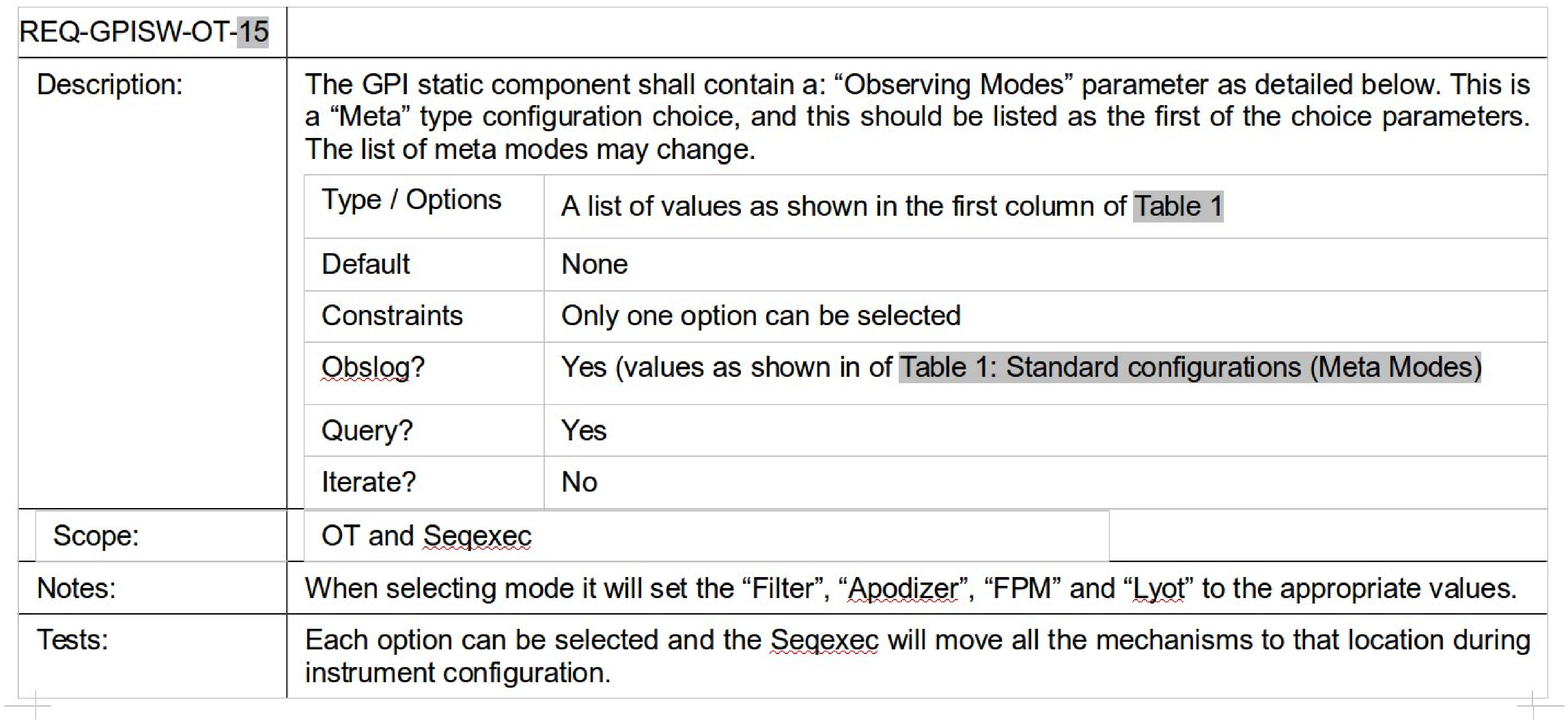}
   \end{tabular}
   \end{center}
   \caption[] 
   { \label{fig:swreq} 
Example of a software requirement. }
   \end{figure} 

Most of the requirements were later broken up into several specific software tasks. The requirements also included a list of all the FITS keywords going into the GPI FITS file headers. This later list required several iterations with the team and the coordinator of the GSA(see section\,\ref{sec:archive}) to have the proper FITS header structure. The final version has 260 unique keywords allowing a complete description of the telescope environment, GPI configuration, GPI performance, and, weather conditions.  
\subsection{Software implementation phase} 
\label{sec:jira}  
The software development inside Gemini uses JIRA as a tool to keep track of tasks, progress and priorities. Each of the requirements defined in section\,\ref{sec:swreq} was broken down if needed into clear separate tasks defined in the JIRA database. 


Each of the defined tasks was ranked in priority/impact and then the software developer estimated the workload to implement the required task and added details on how the task was going to be implemented. The task then passed back to its creator who could then update/clarify the requirement in the task. Some tasks only required a single loop of feedback but several more complex tasks required many iterations to allow the implementation of the task as envisioned by their creator. Tasks were also updated to take into account practical concerns like development time and feasibility of the task description.

In parallel with the predefined tasks any issue/problem found during the installation, science verification and commissioning was taken from the fault reports and converted into separate JIRA tasks.
In total 260 JIRA tasks where created based on requirements, fault reports, bugs and feature improvements. 

GPI integration to the observatory required several changes to the OCS. In particular, the TCC and Seqexec needed several changes to accommodate the instrument characteristics and let it operate along other instruments. GPI’s requirement for an Alignment and Calibration step after every slew, added complexity to the Seqexec as this step has not been needed for other instruments. The ODB was also updated to allow it to store and configure GPI observations and handle the extra calibration steps.

The TCC/TCS was modified to be able to talk to the AO on GPI (in TCC terms this is labelled the {\it On Instrument Wavefront Sensor} (OIWFS), through the GIAPI layer. The most fundamental requirements were that the TCC should be able to open and close all the loops inside GPI when on target and accept offsets from GPI when the loops are closed. Depending on the Zernicke order they are either sent to the secondary mirror (M2) or the primary mirror (M1).  

Phase 1 Tool (PIT), the Observing Tool (OT) and the Sequence Executor (SeqExec) changes/requirements are described further in detail in section\,\ref{sec:obsprep}. 
\section{Observation Preparation}
\label{sec:obsprep}
Gemini observatory are using three main tools for the process from PI proposals to the execution of the defined program. The Phase 1 Tool (PIT), the Observing Tool (OT) and the Sequence Executor (SeqExec) that are described in depth in the following sections.  
\subsection{PIT-Phase 1 Tool}
\label{sec:pit}
The PIT is a GUI written in JAVA used to submit proposals to Gemini Observatory. To offer GPI as an instrument for the Call for Proposals then several updates to the PIT software was made. 
\begin{itemize}
\item Adding GPI as one of the selectable instruments in the resource configuration. 
\item Adding the logic in the instrument configuration so that the proper configurations for GPI can be selected. GPI is a very complex instrument and a lot of work has gone into making the observation definition quick and simple. In the final version of the tool the PI only have to make three choices to fully define the observing mode with GPI: 
\begin{itemize}
\item Select GPI as the instrument
\item Choose the {\it Meta Mode}, choosing between the {\it Coronography}, {\it Direct} Imaging, and, {\it NRM} modes and the desired filter ({\it Y, J, H, K1}, and {\it K2}). 
\item Select the disperser setting, either {\it Spectroscopy} or {\it Polarization}. 
\end{itemize}
   \begin{figure}
   \begin{center}
   \begin{tabular}{c}
   \includegraphics[height=9cm, width=16.5cm]{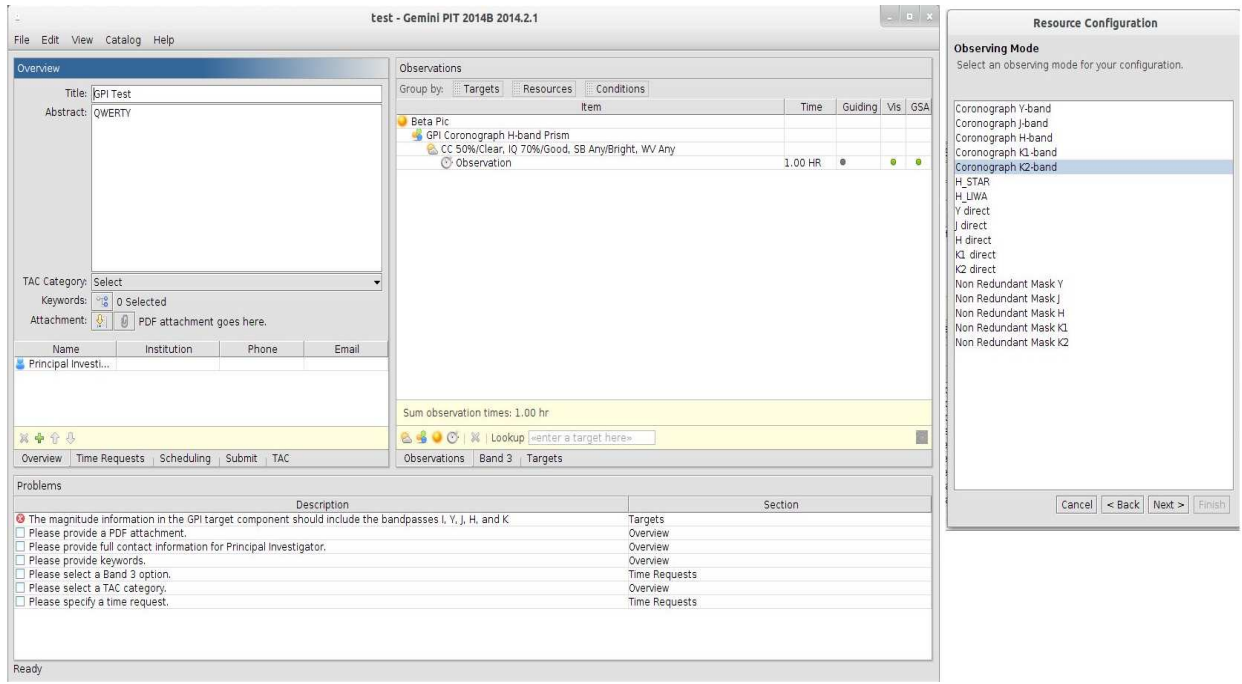}
   \end{tabular}
   \end{center}
   \caption[] 
   { \label{fig:pit} 
Left: The main interface of PIT with example observation defined. Right: Example of the {\it Metamode} choice. }
   \end{figure} 

These three choices allows the PI to completely define the observing configuration for the instrument. 
\item Logic was added to the target component so that the required magnitude information is queried and automatically added to the target field and highlighting any missing information. 
\end{itemize}
The accepted proposals are then imported into the Gemini database and the simple definitions of modes and targets are used to create {\it skeleton templates} that are imported into the {\it Observing Tool}, see the following section. 
\subsection{OT-The Observing Tool}
\label{sec:ot}
In spite of the name of the software this is also used by the PI to define the accepted observations in addition to serving as an interface to the Gemini Observing Data Base (GODB) and choosing the desired observation that is sent to the observing queue. 

One of the complexities at earlier semesters was for the PI to create properly formatted and structured observing sequences. A major step in usability is the use of the skeletons created from the PIT tool. These PIT skeletons and underlying logic for the instrument allows the automatic creation of properly structured observing sequences. The skeletons are visible in the OT in the template folder, and to create the proper observing sequences the PI needs only to click the {\it Apply} button, which will:
\begin{itemize} 
\item Create an observing group for each target with proper labeling. This group contains an {\it Acquisition}, {\it Science} observation and an {\it Arc} if in {\it Spectroscopy} mode. The latter is to allow subpixel corrections to the spectral extraction and the wavelength calibration as there is fractional shifts of the spectra with telescope elevation moves. The arc filter is also set in for the observing wavelength except for  {\it K1} or {\it K2} where exposure times would be too long, where the filter is set to {\it H}. 
\item Parse the selected filter and if {\it K1} or {\it K2} add a sky sequence to the end of the science observation. 
\item Configure the {\it Acquisition} so that it correctly sends the {\it Align\&Cal} sequence that makes the proper internal calibrations after slewing to a new target and/or changing configuration.
\item The logic detects {\it Polarization} observations and properly configures sets of {\it Half-wave} plate angles allowing for proper {\it Stokes} measurements.  
\end{itemize}
Thus with a single click the PI obtains properly structured observing sequences and can then focus on setting the proper exposure times and repeats so that sufficient time per target is done. All of the complexity of configuring the different {\it Apodizers}, {\it Focal} plane masks, {\it Disperser}, {\it Sky Offset} sequences, {\it arcs}, {\it Cassegrain angle}, and, {\it half-wave} plate settings are all properly implemented by the specially defined GPI logic. The only additional choice is flagging a sequence as {\it Astrometric Field}, which is used to set the proper flags for the {\it GPI Pipeline} reduction. 

   \begin{figure}
   \begin{center}
   \begin{tabular}{c}
   \includegraphics[height=9cm]{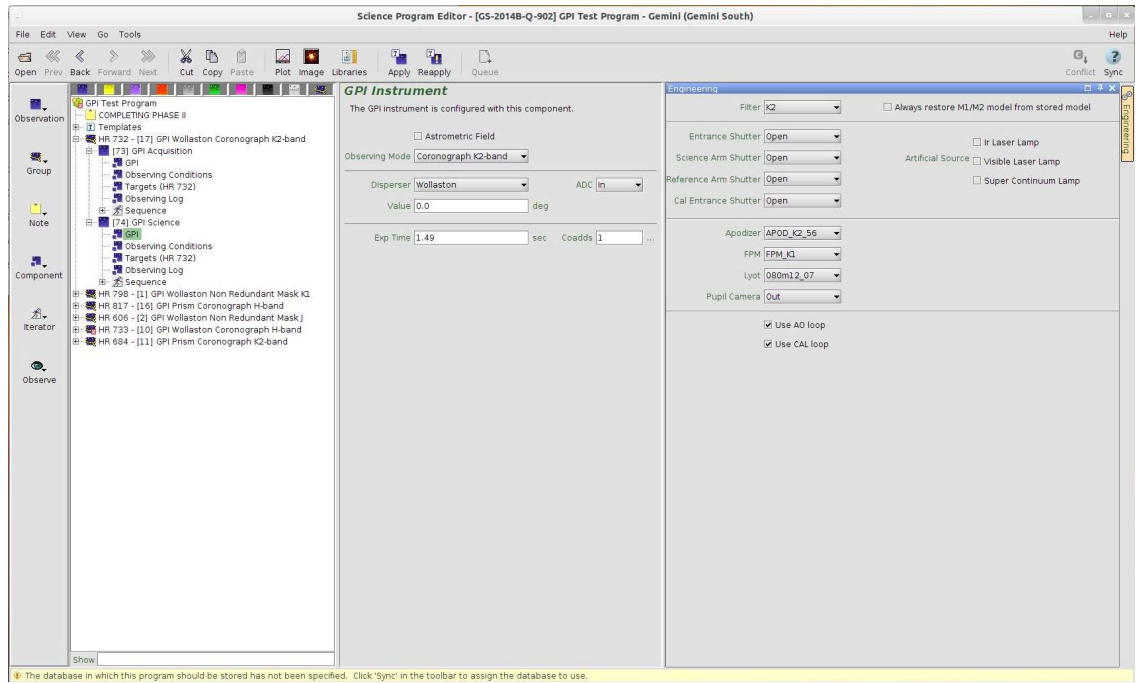}
   \end{tabular}
   \end{center}
   \caption[] 
   { \label{fig:ot} 
The main interface of the OT with example observation defined and the GPI component selected. The {\it Engineering} interface tab is shown where non-standard configurations are available for the expert on-site user, usable for engineering tests.}
   \end{figure} 
\subsubsection{Timeline}
\label{sec:timeline}  
As the Gemini Observatory is based on partner shares in the time allocation then it is fundamental that the timeline as described in the OT (see section\,\ref{sec:ot}) gives proper estimates of the estimated time it takes to take the required science observations including all overheads. With an accurate timeline the astronomer can design the observing sequences and still fit into the allotted time for the program. For GPI it required that the time for component changes is properly determined ($30s$) and the time for exposing and reading out the science data. The default formula for reading out data is:
\begin{itemize}
\item $EXPOSURE + READOUT + DHS$, where $EXPOSURE$ is the total time exposing with the instrument, $READOUT$ is the time reading out the exposure, and $DHS$ the time to properly write the file to the archive directory. 
\end{itemize}   
where for GPI we determined that $READOUT=COADDS\times 3s + 10s$, $EXPOSURE=EXPTIME \times COADDS$, and $DHS=7s$. The $COADDS\times 3s$ overhead are due to physical contraints of the detector, while the other overheads are related to FITS header constrcution and moving of the datafiles to the proper place. With the accurate timeline it makes it possible for an astronomer defining the observing sequences to know the expected time for the execution with high accuracy.   
\subsection{Seqexec-The Sequence Executor}
\label{sec:seqexec}
The observer will highlight the desired program using the {\it OT} and send it to the queue. The queued observation is then loaded into the {\it Seqexec}. The {\it Seqexec} has special logic included so that it parses the {\it xml} definitions that is the output of the {\it OT} and creates a properly structured command sequence that sends commands through the {\it GIAPI} to the {\it TLC}. The tool configures the instrument and controls any exposure sequences set to the instrument. In the case of external calibrations ({\it arcs}) it also configures the GCAL lamps properly. As the tool is a statemachine and has no memory of previous steps a lot of logic had to be added to properly sequence events on both the telescope and instrument side. Thanks to the work on {\it GIAPI} and the {\it TLC}\cite{dunnthis} the needed additional logic could be kept to a minimum.    
\subsection{Acquisition and Science Observation}
\label{sec:acq}  
Although both the telescope environment and the instrument are complex systems, the observations can be broken into two main steps, the {\it Acquisition} and the {\it science sequence}. Thanks to the implemented logic in {\it OT}, {\it Seqexec} and the {\it GIAPI} layer to handle the GPI specific tasks the operation is streamlined and the similar to other instruments used at Gemini allowing quick training of telescope operator and observer the instrument specific details for GPI. The modular structure and the common tools ({\it PIT}, {\it OT}, {\it TCS}, and, {\it Seqexec} between all the instruments enables the complete observation of GPI sequences by two people, the telescope operator and the observer. 

The sequence that describes a complete GPI observation can be broken down into the following steps:  
\begin{itemize}
\item Slew of the telescope to the target position. This is accomplished by loading the queued OT step into the TCS and sending the telescope to the target coordinates. 
\item The error is positioning after the slew is typically of the order of 10\,arcseconds, as the {\it FOV} of GPI is only 2.4\,arcseconds then this is not accurate enough to acquire the object on the GPI {\it AOWFS}. To move the object to the center of the GPI {\it FOV} we use the telescope {\it Acquisition} camera and the object is centered on the pre-determined hot-spot, after which the light is sent back to the instrument, now with a position error of typically 0.5\,arcseconds. The hot-spot coordinates have been carefully measured so that the coordinates corresponds to the center of the {\it FOV} of GPI. Repeated mounting and dismounting of the instrument on the telescope have shown that these coordinates only change by fractions of an arcsecond and this touching up of the hot-spot coordinates are remeasured every time GPI is mounted on the telescope. As GPI also has an optional {\it ADC}\cite{pascale.hibon.spie2014} unit the hot-spot is configured from a look-up table and usage of {\it ADC} or not. This means that the total time of the slew and centering up on the GPI {\it FOV} is of the order of a few minutes. 
\item In parallel with the slew of the telescope the astronomer/observer configures the instrument by loading the sequence into the {\it Seqexec} which configures the instrument in the desired mode.  
\item When the telescope is in position the observer triggers the {\it Align\&Cal} acquisition step that commands the instrument to take the proper internal calibrations\cite{dmitry.savransky.spie2014}. 
\item The final step in the acquisition is the centering and locking of the guide-loops on the {\it AOWFS}. As the object is close to the center of the GPI {\it FOV} the telescope operator reads the measured offsets of the  {\it AOWFS} in the GPI {\it AOWFS} Status Display (see figure\,\ref{fig:ao}) and makes minor telescope pointing offsets so that the offsets seen by the  {\it AOWFS} is less than 100\,milliarcseconds. When centered the operator then triggers the closing of the  {\it AOWFS} and {\it LOWFS} control loops through the {\it TCS}. The {\it TCS} communicates through the {\it GIAPI} interface layer with the instrument and closes the loops. This step completes the so called {\it Acquisition}. 
   \begin{figure}
   \begin{center}
   \begin{tabular}{c}
   \includegraphics[height=10cm]{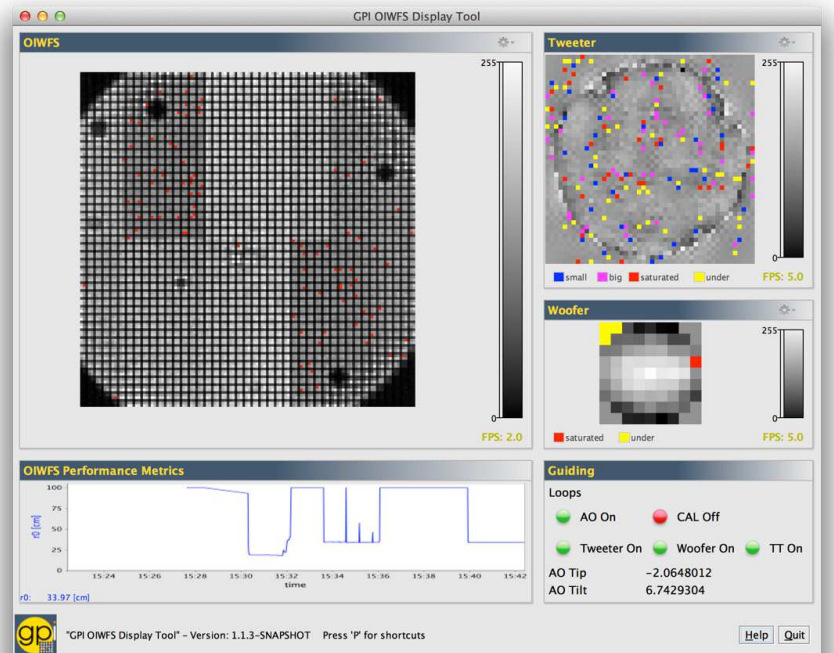}
   \end{tabular}
   \end{center}
   \caption[] 
   { \label{fig:ao} 
The GPI {\it AOWFS} status display. This compact {\it GUI} developed at Gemini gives detailed information on the loop status, {\it AOWFS} performance metrics, offsets, and, tweeter and woofer information.}
   \end{figure} 
\item With the object acquired and loops closed the observer loads in {\it Seqexec} the science sequence and starts the observation. This sequence configures again the instrument for science and executes the predefined science observation sequence. 
\item Science frames are analyzed by the real-time GPI pipeline reduction allowing the observer to evaluate the instrument performance in real-time.      
\item After the science sequence completes the observer loads and executes the {\it arc} sequence that takes a properly configured arc at the current telescope position. This last step finishes the observing sequence and the next target can be chosen and slewed to.   
\end{itemize}

   \begin{figure}
   \begin{center}
   \begin{tabular}{c}
   \includegraphics[height=10cm]{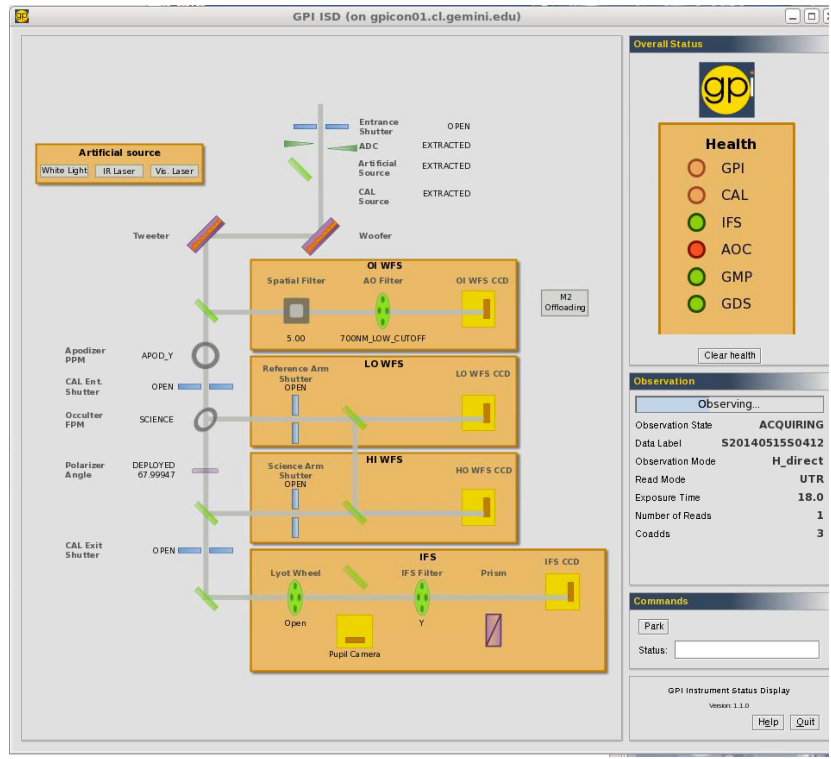}
   \end{tabular}
   \end{center}
   \caption[] 
   { \label{fig:isd} 
The GPI Instrument Status Display ({\it ISD}) is a compact interface showing all the configuration of the instrument, exposure times configuration and progress, and, the instrument status. The display is designed so that observer can quickly see the instrument configuration and GPI system status. This display was designed by Gemini based on the software requirements as detailed in section\,\ref{sec:swreq}}
   \end{figure} 
\section{Data archiving}
\label{sec:archive}  
A fundamental part of the observatory support structure is the Gemini Science Archive (GSA) where all science observations are stored and available for download within a few minutes of the data being taken on sky. 

For our current facility instruments, the Data Handler Software (DHS) merges FITS headers with the raw data and writes the datasets into a directory, then copies that file to the operations storage area. That process is automatic, and gets registered in the Observation Log/Events in the OT for a given observation. In parallel with this process the DataManager software, which is part of the archiving software tools, copies the new raw file to the {\it dataflow} directory and to the transfer area in a separate machine at the base facility. It is from this area that the e-transfer system will pick the new file and copy it to CADC for ingestion in the GSA. When the observer sets the quality assessment flags in the OT, the DataManager software edits the header accordingly and re-transfer the updated file (we call those the ``QA-ed'' datasets). However, the DataManager ONLY monitors the files that are listed in the database, and the creation of an entry in the database comes from the corresponding observing event (seqexec) not by the presence of  the file in the {\it dataflow} directory.

Once the file is in the GSA transfer area, it is picked up by the transfer daemon, it is checked for existence in the Archive (either as completely new or as modified), then it is actually copied to CADC, and finally ``post-processed'' there, when the headers are checked, then parsed into the database tables, and the file is made available to the proper access groups. Once a file is successfully transferred, it is removed from the transfer area.

To allow GPI to properly ingest the files into the archive, the following needed to occur:
\begin{itemize}
\item Create proper FITS header keyword dictionaries, that describes in detail the keywords present in the GPI FITS files. Each of the few hundred keywords was defined with allowed values and if the presence of the keyword is mandatory or not. Absence of Mandatory keywords flags the file as incomplete and it is not transferred to the archive as the absence indicates a major fault in the file structure. Naming of the various keywords were checked to comply with the Gemini naming standard and structure. 
\item The GIAPI software was adjusted so that the files written by the instrument obtained the proper FITS keywords from the telescope environment and the files written to the GSA transfer area. 
\end{itemize}
The instrument is now fully integrated and any science or calibration observations that produces files have the files transferred and ingested into the GSA database. Any registered CADC user can thus access the files through the web-interface and download the files from their resulting database queries. 
\section{Early Science}
\label{sec:earlyscience}
The final test of all the components in the telescope environment, the observation preparation tools, observations and archiving was done in the Early Science ({\it ES}) taking place in mid April 2014. The {\it ES} started with a limited Call for Proposals for observations with the instrument, using the modified {\it PIT}, proposal evaluations and selection of the chosen programs, observation preparations using the {\it OT} and finally observations using the {\it OT}, {\it Seqexec}, and the {\it TCS}. 

The {\it ES} was a great stress test of all the software and processes. The observatory received 30 proposals of which 16 were selected based on their science merit and suitability to test all modes of operation of the instrument. Preparations using the {\it PIT} went smoothly and import and final observation definitions in {\it OT} worked well. The actual observing run lost two nights due to poor weather, but in spite of the challenging conditions 13 of the 16 programs could be completed and 1 program partially completed. 

The {\it ES} process was valuable in allowing further modification of the {\it PIT} and {\it OT} tools, the skeleton handling logic in the {\it OT} so it is expected that the full process will be even smoother for the 14B semester starting in August 2014. The modifications are focusing on stability of observations and easy of use of the software tools for both the PI defining the observations and the actual observers taking the observations in the queue.    

\acknowledgments     
The Gemini Observatory is operated by the Association of Universities for Research in 
Astronomy, Inc., under a cooperative agreement with the NSF on behalf of the Gemini 
partnership: the National Science Foundation (United States), the National Research 
Council (Canada), CONICYT (Chile), the Australian Research Council (Australia), 
Minist\'erio da Ci\'encia, Tecnologia e Inova\c{c}\=ao (Brazil), and Ministerio de Ciencia, 
Tecnolog\'ia e Innovaci\'on Productiva (Argentina).

\bibliography{report}   
\bibliographystyle{spiebib}   
\end{document}